%% file: Jetcas2012_draft.tex
\documentclass[journal]{IEEEtran} 
\input{packages}

\def\x{{\mathbf x}}

\def\y{{\mathbf y}}
\def\z{{\mathbf z}}
\def\w{{\mathbf w}}
\def\m{{\mathbf m}}
\def\n{{\mathbf n}}
\def\h{{\mathbf h}}
\def\v{{\mathbf v}}
\def\p{{\mathbf p}}
\def\s{{\mathbf s}}
\def\d{{\mathbf d}}

\def\L{{\mathbf \Lambda}}
\def\U{{\mathbf U}}
\def\S{{\mathbf S}}
\def\G{{\mathbf G}}
\def\M{{\mathbf M}}
\def\H{{\mathbf H}}
\def\F{{\mathbf F}}

\def\R{{\mathbf R}}
\def\Dt{{\mathbf D_{\mathbf t}}}
\def\I{{\mathbf I}}

\def\Ps{{\mathbf \Psi}}
\DeclareMathOperator*{\argmin}{arg\,min}
\DeclareMathOperator{\TV}{TV}
\DeclareMathOperator{\soft}{\mathfrak{S}}

\begin{document}
%
\title{High Speed Compressed Sensing Reconstruction in Dynamic Parallel MRI Using Augmented Lagrangian and Parallel Processing}

\author{
\IEEEauthorblockN{\c{C}a\u{g}da\c{s}~Bilen, Yao~Wang and Ivan~Selesnick}
\\
\IEEEauthorblockA{Department of Electrical Engineering, 
Polytechnic Institute of NYU, 
Brooklyn, NY, USA}
}


%
%


\maketitle

\begin{abstract}
Magnetic Resonance Imaging (MRI) is one of the fields that the compressed sensing theory is well utilized to reduce the scan time significantly leading to faster imaging or higher resolution images. It has been shown that a small fraction of the overall measurements are sufficient to reconstruct images with the combination of compressed sensing and parallel imaging. Various reconstruction algorithms has been proposed for compressed sensing, among which Augmented Lagrangian based methods have been shown to often perform better than others for many different applications. In this paper, we propose new Augmented Lagrangian based solutions to the compressed sensing reconstruction problem with analysis and synthesis prior formulations. We also propose a computational method which makes use of properties of the sampling pattern to significantly improve the speed of the reconstruction for the proposed algorithms in Cartesian sampled MRI. The proposed algorithms are shown to outperform earlier methods especially for the case of dynamic MRI for which the transfer function tends to be a very large matrix and significantly ill conditioned. It is also demonstrated that the proposed algorithm can be accelerated much further than other methods in case of a parallel implementation with graphics processing units (GPUs).
\end{abstract}

%
\IEEEpeerreviewmaketitle

\section{Introduction}
Compressed sensing and sparse reconstruction methods have been popular topics of research especially in the last decade. Under certain conditions, such as the data being sparse, i.e. with a few non-zero coefficients in a domain that is incoherent with measurement domain, compressed sensing enables reliable recovery of signals even if they are measured at a rate below the Nyquist rate \cite{Candes2008}. This stimulated research in many different fields in which data acquisition is limited due to constraints.

Medical imaging is one of the application areas that adopted compressed sensing principles rather quickly. It is shown that in magnetic resonance imaging (MRI), the number of measurement samples and thus the scan time can be reduced while preserving image quality if compressed sensing principles are used \cite{Lustig2008}. This is further improved by parallel imaging with algorithms such as SparseSENSE utilizing multiple coils \cite{Liu2008}. These initial studies with compressed sensing were on still images or volumes and spatial total variation (TV) or wavelets were used as regularization constraints. Dynamic imaging is shown to benefit from compressed sensing even further due to the images being significantly correlated along temporal dimension and therefore represented with sparse temporal transforms. Similarly to the spatial case, temporal TV and wavelets are commonly used in MRI \cite{Chen2009}, \cite{Gamper2008}. The temporal Fourier transform is also utilized with k-t SparseSENSE \cite{Otazo2010} especially with cardiac MRI, due to the cardiac motion being periodic therefore sparse with respect to Fourier transform. K-t Group Sparse SENSE is introduced by Usman et al. \cite{Usman2011} which groups pixels with respect to their spatio temporal activity to treat static and dynamic regions differently during reconstruction. 

Parallel computing has gained a great deal of interest in recent years and processors with multiple cores has become a standard in commercial computing. With the introduction of NVidia CUDA and ATI Stream architectures, it is possible to make use of large number of processor cores in graphics processing units (GPUs) for general purpose computing. Image reconstruction in the field of MRI has been shown to benefit from parallel processing and GPUs for real-time reconstruction \cite{Soransen2008}, \cite{Soransen2009}. Regardless of the algorithm used, the reconstruction in compressed sensing MRI takes significantly more time compared to reconstruction of regular MRI. However real-time or fast reconstruction is essential in order for the commercial implementations of compressed sensing MRI to be useful in the field of medicine. Therefore fast reconstruction algorithms that can be implemented in parallel are necessary for compressed sensing MRI to be commercially viable.

In all the compressed sensing based approaches to medical imaging, the images are reconstructed by enforcing sparsity of the signal in a separate transform domain while simultaneously having the constraint of consistency with the measurements. This is often formulated as a basis pursuit problem and many different algorithms has been proposed for the solution such as iterative shrinkage based methods (FISTA \cite{Beck2009}, TwIST \cite{Bioucas2007}). The Augmented Lagrangian based methods or mainly Alternating Direction Method of Multipliers (ADMM) approach \cite{Boyd2011} which is mostly equivalent to the Split-Bregman Method \cite{Goldstein2009} has gained significant popularity especially in recent years due to its flexibility, ease of implementation and speed. SALSA algorithm which is also based on ADMM, has been shown to converge faster than other popular methods in various different imaging inverse problems \cite{Afonso2010a} combining the ideas in ADMM with other denoising methods. Parallel MRI however suffers from slower implementations due to the transfer function being large as well as harder to adapt to ADMM formulation efficiently. Recently ADMM approach has been formulated for parallel MRI reconstruction \cite{Ramani2011}, which mainly deals with still image reconstruction with spatial regularization, however the speed of the presented algorithms has not been evaluated for dynamic MRI reconstruction with temporal regularization, which lead to very large and severely ill-conditioned transfer functions.

In this paper, we present two ADMM based solutions to analysis and synthesis prior formulations for compressed sensing dynamic parallel MRI reconstruction and propose an efficient implementation of the transfer function for the proposed ADMM algorithms in order to aid significantly faster reconstruction. The proposed method exploits the commonly used Cartesian subsampling pattern in MRI to provide faster and efficient implementation. It also benefits more from a parallel implementation such as in GPUs. It is demonstrated by the experiments that the proposed algorithm is faster and can be accelerated further than earlier methods when implemented with GPUs. 

\section{Problem Formulation} 
\subsection{Compressed Sensing and Parallel MRI}
In magnetic resonance imaging, the measurements related to the image of a 2D slice of a volume are acquired in the spatial Fourier transform domain, which can be represented as
\begin{align}
\label{eqn:mri_basic}
\y_{t} &= \F\x_{t} + \n \\
&= \F_\h\F_\v\x_{t} + \n
\end{align} 
where $\F_\h$ and $\F_\v$ are matrices representing Fourier transform operation along horizontal and vertical axes and $\F$ represents the 2D spatial Fourier transform. $\x_{t}$ and $\y_{t}$ represent image of the 2D slice and corresponding Fourier measurements at time $t$ respectively and $\n$ represents the random noise in the measured samples that is modelled as Gaussian. The equation in  (\ref{eqn:mri_basic}) can easily be expanded for 3D volumes instead of slices, but all the derivations in this paper will consider 2D slices for simplicity. The vector $\x_{t}$ is defined as
\begin{equation}
\label{eqn:x_def}
\x_{t} \triangleq \left[
\begin{array}{c}
\x_{\v,1,t} \\ \vdots \\ \x_{\v,n_h,t} 
\end{array}
\right], 
\x_{\v,i,t} \triangleq \left[
\begin{array}{c}
x_{1,i,t} \\ \vdots \\ x_{n_v,i,t} 
\end{array}
\right]
\end{equation} 
where $x_{j,i,t}$ is the image pixel at horizontal position $i$, vertical position $j$ and at time $t$ of an image of $n_h \times n_v$ spatial resolution and $n_t$ frames. 

Parallel MRI uses multiple receiving coils to speed up the data acquisition by making use of the redundancy among different coils.  The measurement made by $c$-th coil can be represented as
\begin{align}
\label{eqn:mri_mcoil}
\y_{t,c}& = \F\S_c\x_{t} + \n\\
\label{eqn:mri_mcoil2}
& = \F_\h\F_\v(\s_c\odot\x_{t}) + \n \\
\nonumber 
\s_c &\triangleq \left[
\begin{array}{c}
\s_{\v,1,c} \\ \vdots \\ \s_{\v,n_h,c} 
\end{array}
\right], \s_{\v,i,c} \triangleq \left[
\begin{array}{c}
s_{1,i,c} \\ \vdots \\ s_{n_v,i,c} 
\end{array}
\right]
\end{align} 
where $s_{j,i,c}$ indicates the sensitivity or weight associated with pixel $x_{j,i,t}$ by coil $c$. The matrix $\S_c$ is a diagonal matrix that has the vector $\s_c$ along its diagonal, while $\odot$ represents element by element multiplication of two vectors. The sensitivity matrices are acquired after calibration and often noise decorrelation so that $\n$ has a Gaussian distribution. 

\begin{figure}
\centering
\subfloat[][\label{fig:mask2d}]{
\includegraphics[scale=1]{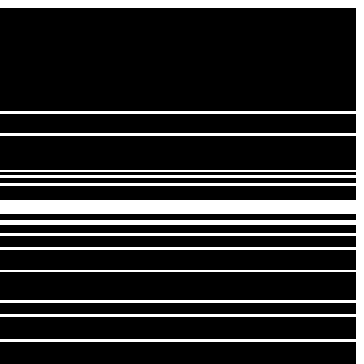}
}%
\qquad
\subfloat[][\label{fig:mask3d}]{
\includegraphics[scale=1]{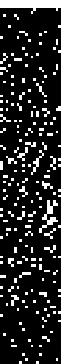}
}
\caption{A subsampling pattern example in dynamic MRI where only $1/8$ of the samples are scanned (represented by white points) (a) Subsampling pattern for single frame in spatial Fourier domain (b) Sampled vertical indices along time in Dynamic MRI}%
\label{fig:Subsampling}%
\end{figure}

In compressed sensing, only a subset of the measurements (often selected randomly) are acquired therefore reducing the required scan time in MRI. In Cartesian sampling, all the Fourier coefficients along the horizontal axis corresponding a specific vertical coordinate are measured sequentially and the samples cannot be skipped to save the acquisition time. Each new horizontal line however requires a switching of magnetic field gradient, which has a minimal switching time limit and the total time spent for acquisition is proportional to the number of scanned vertical lines, therefore the subsampling is usually only along vertical axis in the Fourier domain. In dynamic MRI, the subsampled vertical coordinates are also changed in time randomly. This process can be represented for each coil as
\begin{align}
\label{eqn:mri_csmcoill}
\tilde{\y}_{t,c} &= \H_{t,c}\x_{t} + \n\\
\label{eqn:mri_csmcoill2}
 &= \F_\h\M_{\v,t}\F_\v\S_c\x_{t} + \n\\
 &= \F_\h\tilde{\F}_{\v,t}\S_c\x_{t} +\n\\
\label{eqn:H_2}
\H_{t,c} &\triangleq \F_\h\tilde{\F}_{\v,t}\S_c
\end{align} 
where $\tilde{\y}_{t,c}$ is the vector of the sampled subset of measurements for coil $c$ at time $t$ and $\tilde{\F}_{\v,t} = \M_{\v,t}\F_\v$ is the subsampled Fourier transform along the vertical axis which is composed of subsampled Fourier Transform matrices $\tilde{\F}_{\v,i,t}$ transforming along each horizontal coordinate $i$ 
\begin{equation}
\tilde{\F}_{\v,t} = \left[ \begin{array}{ccc}
\tilde{\F}_{\v,1,t} & \cdots & 0 \\
\vdots & \ddots & \vdots\\
0 & \cdots & \tilde{\F}_{\v,n_h,t}
\end{array} \right]
\end{equation}
$\F_\h$ in (\ref{eqn:mri_csmcoill2}) represents the Fourier transform along the horizontal axis similar to (\ref{eqn:mri_mcoil2}), but the size of the matrix might be different due to subsampling. Other subsampling patterns are also possible using different methods of scanning \cite{Lustig2008}. A sample subsampling pattern for dynamic MRI can be seen in Figure~\ref{fig:Subsampling}.

The transfer function in dynamic MRI with parallel coils can be represented as
\begin{align}
\y &= \tilde{\F}\S\x +\n\\
\label{eqn:mri_cspmril}
	&=\H\x + \n 
\end{align}
where $\y$, $\x$, $\tilde{\F}$, $\S$ and $\H$ can be defined as
\begin{equation}
\x \triangleq \left[
\begin{array}{c}
\x_{1} \\ \vdots \\ \x_{n_t} 
\end{array} \right], 
\y \triangleq \left[
\begin{array}{c}
\y_{1} \\ \vdots \\ \y_{n_t} 
\end{array} \right], 
\y_{t} \triangleq \left[
\begin{array}{c}
\tilde{\y}_{t,1} \\ \vdots \\ \tilde{\y}_{t,n_c} 
\end{array} \right] 
\end{equation}
\begin{align}
\label{eqn:H_1}
\H &\triangleq \left[ \begin{array}{ccc}
\H_1 & \cdots & 0 \\
\vdots & \ddots & \vdots\\
0 & \cdots & \H_{n_t}
\end{array} \right], 
\H_t \triangleq \left[ \begin{array}{c}
\H_{t,1} \\
\vdots\\
 \H_{t,n_c}
\end{array} \right] \\
\label{eqn:S}
 \S &\triangleq  \left[\begin{array}{ccc}
 \S^* & \cdots & 0\\ 
 \vdots & \ddots & \vdots \\
 0 & \cdots & \S^*
\end{array}
\right],
\S^* \triangleq  \left[\begin{array}{c}
 \S_{1}  \\ \vdots \\ \S_{n_c}
\end{array}
\right],\\
\label{eqn:F}
\tilde{\F}  &\triangleq\left[\begin{array}{ccc}
 \F^*_1 & \cdots & 0\\ 
 \vdots & \ddots & \vdots \\
 0 & \cdots & \F^*_{n_t}
\end{array}
\right]
\F^*_t \triangleq \left[\begin{array}{ccc}
 \F_\h \tilde{\F}_{\v,t} & \cdots & 0\\ 
 \vdots & \ddots & \vdots \\
 0 & \cdots & \F_\h \tilde{\F}_{\v,t}
\end{array}
\right]
\end{align} 
and satisfy $\H=\tilde{\F} \S$. $\S$ and $\F^*_t $ are block diagonal matrices that are composed of $n_t$ and $n_c$ blocks along the diagonal respectively.

Compressed sensing theory shows that the signal $\x$ can be recovered from $\y$ almost certainly if,
\begin{enumerate}
\item $\Ps\x$ is sufficiently sparse for a known transform $\Ps$ ($\Ps'\Ps=\Ps\Ps'=\I$, $.'$ indicating the conjugate transpose)
\item $\H$ and $\Ps$ are sufficiently incoherent, i.e. the off diagonal entries of $\Ps\H'\H\Ps'$ after normalization are sufficiently small
\end{enumerate}
One of the breakthroughs in compressed sensing is that the second condition is shown to be satisfied when $\H$ is formed by random entries, or by a random subset of rows of a full rank matrix as in (\ref{eqn:mri_csmcoill}) regardless of $\Ps$ \cite{Candes2008}, \cite{Lustig2008}. The first condition is also relaxed such that $\Ps$ can be an overcomplete transform or dictionary or a penalty operator as in case of total variation \cite{Rauhut2008}. Under these conditions it is shown that $\x$ can be recovered almost certainly with the minimization 
\begin{align}
\label{eqn:L0}
\tilde{\w} &= \argmin_{\w} \Vert\w\Vert_0 \text{   s.t.  } \Vert \y - \H\Ps'\w \Vert^2_2 \leq \epsilon \\
\tilde{\x} &= \Ps'\tilde{\w}
\end{align}
in which $\Vert.\Vert_0$ is the $L_0$-norm (which is in fact a quasi-norm) that is the number of non-zero entries of a vector ($\Vert \x \Vert_p \triangleq \left(\sum\limits_i x_i^p\right)^{1/p} $ for $p>0$). The constraint $\Vert \y - \H\Ps'\w \Vert^2_2 \leq  \epsilon$ ensures consistency with the measurements and it is optimum for i.i.d. Gaussian noise with standard deviation $\epsilon$. The minimization in  (\ref{eqn:L0}) is non-convex and therefore practical methods do not guarantee convergence to global minimum. It is shown in the sparse reconstruction literature that the minimization of $L_1$-norm as in
\begin{align}
\label{eqn:L1}
\tilde{\w} &= \argmin_{\w} \Vert\w\Vert_1 \text{   s.t.  } \Vert \y - \H\Ps'\w \Vert^2_2 \leq  \epsilon \\
\tilde{\x} &= \Ps'\tilde{\w}
\end{align}
will lead to the same solution as in (\ref{eqn:L0}) provided that $\w$ is sufficiently sparse \cite{Donoho2009}. The implications of this equivalence is significant since (\ref{eqn:L1}) is a convex optimization problem with a global minimum and can be solved very efficiently with methods such as Alternating Direction Method of Multipliers (ADMM) \cite{Afonso2011} or the Split-Bregman Method \cite{Goldstein2009}. In order to further simplify the minimization, the constraint in  (\ref{eqn:L1}) can be removed to form the unconstrained formulation
\begin{align}
\label{eqn:L1unconst}
\tilde{\w} &= \argmin_{\w} \lambda \Vert\w\Vert_1 + \Vert \y - \H\Ps'\w \Vert^2_2 \\
\tilde{\x} &= \Ps'\tilde{\w}
\end{align}
A number of fast minimization algorithms exist for the solution of unconstrained minimization in (\ref{eqn:L1unconst}) such as TwIST, FISTA and SALSA (\cite{Beck2009}, \cite{Afonso2010a} and references therein) and the result is equivalent to (\ref{eqn:L1}) if $\lambda$ is adjusted properly. 

The formulation in (\ref{eqn:L1}) is referred to as synthesis prior formulation and (\ref{eqn:L1unconst}) can be used with overcomplete transforms such as wavelets or non-orthogonal dictionaries. An alternative approach uses an analysis prior formulation with a penalty term for regularization as in
\begin{equation}
\label{eqn:L1an_pri}
\tilde{\x} = \argmin_{\x} \lambda R(\x) + \Vert \y - \H\x \Vert^2_2
\end{equation}
in which $R(\x)$ is the penalty function that is large when $\x$ has characteristics different from a prior knowledge of $\x$. A commonly used example for such penalty functions is the total variation (TV) which is defined for 1D signals as
\begin{equation}
\TV(\x) \triangleq \sum\limits_i |x_i - x_{i-1}|
\end{equation}
and penalizes the signal gradient. In case of dynamic MRI, temporal regularization is often employed and TV can be defined along the temporal axis as
\begin{align}
\label{eqn:TV}
\nonumber
\TV_t(\x) &= \sum_{t=2}^{n_t} \Vert\x_t - \x_{t-1}\Vert_1 \\
\nonumber &=  
\left\Vert
\begin{bmatrix} 
-\I & \I & 0 & \cdots & 0 \\
0 & -\I & \I & \cdots & 0 \\
\vdots & & \ddots & & \vdots \\
0 & \cdots & 0 & -\I & \I
\end{bmatrix}
\left[ \begin{array}{c} \x_1 \\ \vdots \\ \x_{n_t} \end{array} \right] 
\right\Vert_1 \\
&= \Vert\Dt\x\Vert_1
\end{align} 
in which $\Dt$ is the temporal difference matrix. Minimization of (\ref{eqn:L1an_pri}) with $R(\x)=\TV_t(\x)$ is possible with algorithms such as MFISTA \cite{Beck2009} or ADMM based methods \cite{Goldstein2009}, \cite{Afonso2010a}.

\subsection{ADMM Basics}

Alternating Direction Method of Multipliers (ADMM) is fast convex minimization algorithm that makes use of variable splitting and Augmented Lagrangian to solve various forms of objective functions. For a general minimization problem such as
\begin{equation}
\label{eqn:admm_basic}
\argmin\limits_\z f(\z) = f_1(\z) + f_2(\G\z)
\end{equation}
which may be non-trivial to directly solve for, ADMM algorithm solves the equivalent problem
\begin{align}
\label{eqn:admm_basic2}
\argmin\limits_{\z,\v}& f(\z,\v) = f_1(\z) + f_2(\v) \\
\nonumber &\text{ such that } \G\z = \v
\end{align}
Solving for (\ref{eqn:admm_basic2}) can be much easier than solving for (\ref{eqn:admm_basic}) if functions $f_1$, $f_2$ and the auxiliary variable $\v$ are selected properly. The ADMM algorithm that solves for (\ref{eqn:admm_basic2}) can be summarized as iteratively applying
\begin{align}
\label{eqn:admm_ex1}
\z &\gets \argmin\limits_{\z} f_1(\z) + \mu \Vert \G\z - \v - \d \Vert_2^2 \\
\v &\gets \argmin\limits_{\v} f_2(\v) + \mu \Vert \G\z - \v - \d \Vert_2^2 \\
\label{eqn:admm_ex3}
\d &\gets \d - (\G\z - \v)
\end{align}
until convergence after initializing $\v$, $\d$ and $\mu$ \cite{Boyd2011}. The iterative steps (\ref{eqn:admm_ex1})-(\ref{eqn:admm_ex3}) solving (\ref{eqn:admm_basic2}) is also known as the Split-Bregman Method, and is guaranteed to converge given $\mu>0$, however convergence can be very slow unless $\mu$ is selected properly \cite{Goldstein2009}. The parameter $\mu$ can be selected empirically for a given problem but more discussion on how to select $\mu$ theoretically can be found in \cite{Goldstein2009}, \cite{Boyd2011} and the references therein.

\section{Related Algorithms}
\label{sec:related_alg}
The ADMM approach has been proposed to solve problems of the form (\ref{eqn:L1an_pri}) before, such as the Split-Bregman Method proposed in \cite{Goldstein2009} minimizing
\begin{align}
\label{eqn:anL1}
\tilde{\x} =& \argmin_{\x} \lambda \Vert\v\Vert_1 + \Vert \y - \H\x \Vert^2_2 \\
\nonumber &\text{ such that } \v = \R\x 
\end{align}
assuming the penalty function is in the form $R(\x)=\Vert\R\x\Vert_1$. This formulation is also proposed in \cite{Ramani2011} (named as P1) for Parallel MRI and can be solved iteratively with the steps similar to (\ref{eqn:admm_ex1})-(\ref{eqn:admm_ex3})
\begin{align}
\nonumber \v \gets &\argmin\limits_{\v} \lambda \Vert \v \Vert_1 + \mu \Vert \R\x - \v - \d \Vert_2^2 \\
\label{eqn:admm_bregman_1}
 & =  \soft(\R\x - \d, \dfrac{\lambda}{2\mu}) \\
\nonumber \x \gets &\argmin\limits_{\x} \Vert \y - \H\x \Vert^2_2 + \mu \Vert \R\x - \v - \d \Vert_2^2 \\
\label{eqn:admm_bregman_2}
 &=  (\mu\R'\R + \H'\H)^{-1}\left[ \H'\y + \mu\R'(\v+\d) \right] \\
\label{eqn:admm_bregman_3}
\d\gets  & \d - (\R\x - \v)
\end{align}
where $\soft(.)$ is the element-wise soft thresolding function defined as
\begin{equation}
\soft(a, \tau) \triangleq \left\lbrace \begin{array}{ r }  (|a|-\tau)\dfrac{a}{|a|} \text{ if } |a| > \tau \\ 0 \text{ if } |a| \leq \tau \end{array}\right.
\end{equation}
Note that (\ref{eqn:admm_bregman_2}) requires $(\mu\R'\R + \H'\H)^{-1}$ which is not invertible unless $\R$ and $\H$ have disjoint null spaces. In the case that it is invertible, a closed form solution is often not available especially if $\R'\R\neq\I$ and $\H'\H\neq\I$. For the Parallel MRI with a large $\H$, it is suggested in \cite{Ramani2011}  that the term in (\ref{eqn:admm_bregman_2}) be calculated with few steps of Conjugate Gradient (CG) algorithm at each iteration of P1. However the use of CG iterations is not preferable due to being slow and not precise and can greatly slow down reconstruction when $\H$ or $\R$ is very large. 

SALSA proposed in \cite{Afonso2010a} uses an alternative approach and minimizes
\begin{align}
\label{eqn:anL1_salsa}
\tilde{\x} =& \argmin_{\x} \lambda \Vert\R\v\Vert_1 + \Vert \y - \H\x \Vert^2_2 \\
\nonumber &\text{ such that } \v = \x 
\end{align}
assuming an efficient method exists to solve 
\begin{equation}
\label{eqn:proximal}
\v \gets \argmin\limits_\v\lambda \Vert\R\v\Vert_1+\mu \Vert\v-\x-\d\Vert_2^2
\end{equation}
which sometimes might be another iterative algorithm such as Chambolle's Method used for TV regularization \cite{Chambolle2004}. SALSA replaces $(\mu\R'\R + \H'\H)^{-1}$ in (\ref{eqn:admm_bregman_2}) with $(\mu\I + \H'\H)^{-1}$ which is easier to compute and has a closed form solution for a variety of applications, but still often relies on iterative algorithms to solve for (\ref{eqn:proximal}). Although SALSA has been shown to perform much better than popular methods such as FISTA, MFISTA or TwIST, it can be very slow if  $(\mu\I + \H'\H)^{-1}$ cannot be computed efficiently \cite{Afonso2010a}.

In addition to the algorithms summarized above which have been suggested for generic problems, a second algorithm is proposed in \cite{Ramani2011} (named as P2) specifically for Parallel MRI problem proposing more number of auxiliary variables such that
\begin{align}
\label{eqn:anL1_3}
\tilde{\x} = &\argmin_{\x} \lambda \Vert\v\Vert_1 + \Vert \y - \tilde{\F}\p \Vert^2_2 \\
\nonumber &\text{ such that } \v = \R\m,	\m = \x, \p= \S\x 
\end{align}
where $\tilde{\F}$ and $\S$ are subsampled Fourier transform and sensitivity matrices respectively as defined in (\ref{eqn:S}) and (\ref{eqn:F}) and satisfy $\H = \tilde{\F}\S$. The ADMM algorithm solving (\ref{eqn:anL1_3}) requires calculation of the terms $(\mu\I + \R'\R)^{-1}$, $(\mu\I + \tilde{\F}'\tilde{\F})^{-1}$ and $(\mu\I + \S'\S)^{-1}$ each of which has a closed form solution and can be calculated fast and accurately at each iteration. However although each iteration of the algorithm solving (\ref{eqn:anL1_3}) is quite fast, the number of iterations needed for convergence is much larger than (\ref{eqn:anL1}) or (\ref{eqn:anL1_salsa}) due to having too many auxiliary variables, and the reported speed is comparable to or worse than (\ref{eqn:anL1}) with CG iterations \cite{Ramani2011}.

\section{Proposed Methods}

Considering the drawbacks of the algorithms mentioned in Section~\ref{sec:related_alg}, we propose two algorithms to solve synthesis and analysis prior formulations respectively, which do not depend on any iterative steps except the iterations of ADMM and make use of the fast computation of $(\mu\I + \H'\H)^{-1}$ proposed in Section~\ref{sec:fastHconjH}.

\subsection{ADMM for Synthesis Prior}
Basic variable splitting approach in (\ref{eqn:admm_basic2}) can be used to solve (\ref{eqn:L1unconst}) by
\begin{align}
\label{eqn:syntL1}
\tilde{\w} =& \argmin_{\w} \lambda \Vert\v\Vert_1 + \Vert \y - \H\Ps'\w \Vert^2_2 \\
\nonumber &\text{ such that } \v = \w \\
\tilde{\x} =& \Ps'\tilde{\w}
\end{align}
which in turn can be minimized with iterations similar to (\ref{eqn:admm_bregman_1})-(\ref{eqn:admm_bregman_3}) except $\x$ is replaced with $\w$, $\R$ with $\I$ and $\H$ with $\H\Ps'$. The synthesis prior formulation is often considered as the same problem with (\ref{eqn:anL1}) or (\ref{eqn:anL1_salsa}) with a different transfer function $\H\Ps'$, and can be solved with the same algorithms. However due to the change of transfer function, $(\mu\I + \Ps\H'\H\Ps')^{-1}$ is needed instead of $(\mu\I + \H'\H)^{-1}$ which does not have any closed form solution for an arbitrary $\Ps$ and can be tedious to compute numerically. Assuming $\Ps$ is a tight frame satisfying $\Ps'\Ps=\I$, we propose to use 
\begin{align}
(\mu\I + \Ps\H'\H\Ps')^{-1} = &\dfrac{1}{\mu}\I-\Ps\H'(\mu\I+\H\H')^{-1}\H\Ps'\\
											= & \dfrac{1}{\mu}\I-\Ps(\dfrac{1}{\mu}\I-(\mu\I+\H'\H)^{-1})\Ps'\\
											= & \dfrac{1}{\mu}\I-\dfrac{1}{\mu}\Ps\Ps'+\Ps(\mu\I + \H'\H)^{-1}\Ps'
\end{align}
by making use of the matrix inversion lemma. Therefore the overall speed of the algorithm depends on the fast computation of $(\mu\I + \H'\H)^{-1}$. The algorithm solving (\ref{eqn:L1unconst}) with ADMM with $\Ps$ being a tight frame is summarized in Algorithm~\ref{alg:admm_synth}. Note that the term $\left(\dfrac{1}{\mu}\I-\dfrac{1}{\mu}\Ps\Ps'\right)$ vanish when $\Ps$ is an orthonormal transform and the analysis and synthesis prior formulations become equivalent as are the steps of the algorithms solving (\ref{eqn:anL1}), (\ref{eqn:anL1_salsa}) and (\ref{eqn:syntL1}). But when an overcomplete transform satisfying  $\Ps'\Ps=\I$ is used and a fast computation of $(\mu\I + \H'\H)^{-1}$ is available, Algorithm~\ref{alg:admm_synth} can be much faster than synthesis prior minimization algorithms such as FISTA, or other ADMM based algorithms such as SALSA that uses $(\mu\I + \Ps\H'\H\Ps')^{-1}$.

\begin{algorithm} [!t]
\caption{ADMM minimization to solve (\ref{eqn:L1unconst})}
\label{alg:admm_synth} 
\begin{algorithmic}[1] 
\Procedure{\textit{ADMM\_Synthesis}}{$\y,\H,\Ps,\lambda$}
\Statex \textit{Minimizes $\lambda\Vert\v\Vert_1 + \Vert \y - \H\Ps'\w \Vert^2_2$ s.t $\w=\v,\x = \Ps'\w$}
\Statex
\State Set $\mu>0$, $\d=0, \w=\Ps\H'\y$
\While{\textit{convergence criteria not met}}
	\State 
	$\v \gets \argmin\limits_{\v} \lambda \Vert \v \Vert_1 + \mu \Vert \w - \v - \d \Vert_2^2 $
	\Statex \hspace{0.2\linewidth} $ =	\soft(\w - \d, \dfrac{\lambda}{2\mu})  $\label{alg:minv1}
	 
	\State $\w \gets \argmin\limits_{\w} \Vert \y - \H\Ps'\w \Vert^2_2 + \mu \Vert \w - \v - \d \Vert_2^2$ \label{alg:minx1}
	 \Statex \hspace{0.2\linewidth} $ \begin{array}{l}
	 =\dfrac{1}{\mu}(\I-\Ps\Ps')\left[ \Ps\H'\y + \mu(\v+\d) \right] \\
	 +\Ps(\mu\I + \H'\H)^{-1}\left[ \H'\y + \mu\Ps'(\v+\d) \right]
	 \end{array} $
	 \Statex
	 \State $\d\gets  \d - (\w - \v)$
\EndWhile
\State \textbf{return} $\x = \Ps'\tilde{\w}$
\EndProcedure 
\end{algorithmic} 
\end{algorithm}

\subsection{ADMM for Analysis Prior}
A more challenging problem is the analysis prior formulation in (\ref{eqn:L1an_pri}) which can be approached similar to synthesis prior formulation to minimize 
\begin{align}
\label{eqn:anL1_2}
\tilde{\x} =& \argmin_{\x} \lambda \Vert\v\Vert_1 + \Vert \y - \H\x \Vert^2_2 \\
\nonumber &\text{ such that }  \v = \R\m, \m = \x 
\end{align}
Algorithm solving (\ref{eqn:anL1_2}) is presented in Algorithm~\ref{alg:admm_anlys} and can be derived the same way as Algorithm~\ref{alg:admm_synth}. The auxiliary variable $\m$ in (\ref{eqn:anL1_2}) essentially decouples the regularization and data fidelity as can be seen in lines \ref{alg:minm2} and \ref{alg:minx2} of Algorithm~\ref{alg:admm_anlys}, and the inversions $\left( \frac{\mu_2}{\mu_1} \I + \R'\R\right)^{-1}$ and $(\mu_2\I + \H'\H)^{-1}$ are guaranteed to exist. 

\begin{algorithm} [!t]
\caption{ADMM minimization to solve (\ref{eqn:L1an_pri})}
\label{alg:admm_anlys} 
\begin{algorithmic}[1] 
\Procedure{\textit{ADMM\_Analysis}}{$\y,\H,\R,\lambda$}
\Statex \textit{Minimizes $\lambda\Vert\v\Vert_1+ \Vert \y - \H\x \Vert^2_2$ s.t $\v = \R\m$, $\m=\x$}
\Statex
\State Set $\mu_1,\mu_2 > 0$, $\d_1,\d_2=0, \x=\H'\y, \m=\x$
\While{\textit{convergence criteria not met}}
	\State 
	$\v \gets  \argmin\limits_{\v} \lambda \Vert\v\Vert_1 + \mu_1\Vert\v - \R\m - \d_1\Vert_2^2 $
	\Statex \hspace{0.2\linewidth} $ =	\soft( \R\m + \d_1, \frac{\lambda}{2\mu_1})  $\label{alg:minv2}

	\State $\m \gets \argmin\limits_{\m}  \begin{array}{c}
																	\\
																	 \mu_1\Vert\v- \R\m - \d_1\Vert_2^2 \\
																	  + \mu_2\Vert \m - \x - \d_2 \Vert_2^2
																\end{array}	 $ \label{alg:minm2}
	\Statex \hspace{0.2\linewidth} $ = \left( \frac{\mu_2}{\mu_1} \I + \R'\R\right)^{-1} 
			\begin{array}{l} \\
			 \left[ \R'(\v-\d_1)  \right. \\
			  \left. + \frac{\mu_2}{\mu_1} (\x+\d_2) \right]
			  \end{array} $
	\Statex
	
	\State $\x \gets \argmin\limits_{\x} \Vert \y - \H\x \Vert^2_2 + \mu_2 \Vert \m - \x - \d_2 \Vert_2^2$ \label{alg:minx2}
	 \Statex \hspace{0.2\linewidth} $ =(\mu_2\I + \H'\H)^{-1}\left[ \H'\y + \mu_2(\m-\d_2) \right]$
	 \Statex
	 \State $\d_1\gets  \d_1 - (\v - \R\m)$	 
	 \State $\d_2\gets  \d_2 - (\m - \x)$
\EndWhile
\State \textbf{return} $\x$
\EndProcedure 
\end{algorithmic} 
\end{algorithm}

In order for each iteration of Algorithm~\ref{alg:admm_anlys} to be fast, the terms $\left( \frac{\mu_2}{\mu_1} \I + \R'\R\right)^{-1}$ (line~\ref{alg:minm2}) and $(\mu_2\I + \H'\H)^{-1}$ (line~\ref{alg:minx2}) must be calculated efficiently and if possible precisely. Closed form solution to $\left( \frac{\mu_2}{\mu_1} \I + \R'\R\right)^{-1}$ exist for many common $\R$ and it can otherwise be calculated with few Conjugate Gradient (CG) iterations due to $\R'\R$ often having only few zero eigenvalues. Therefore the computation speed of $(\mu_2\I + \H'\H)^{-1}$ is the main factor determining the speed of each iteration of Algorithm~\ref{alg:admm_anlys}. Note that when compared to traditional ADMM approach in (\ref{eqn:anL1}), Algorithm~\ref{alg:admm_anlys} can be much faster when $(\mu_2\I + \H'\H)^{-1}$ can be quickly computed or has a closed form solution unlike $(\mu\R'\R + \H'\H)^{-1}$, but it will be slower when $\R'\R=\I$ due to the extra variable separation. When compared to SALSA, the extra variable separation decoupling the data fidelity and regularization also provides advantage for regularization terms such as TV, removing the need for other iterative algorithms for TV minimization. The iterations of Algorithm~\ref{alg:admm_anlys} will also converge faster than P2 algorithm in \cite{Ramani2011} due to having fewer auxiliary variables without introducing additional computational complexity assuming computation of $(\mu_2\I + \H'\H)^{-1}$ can be fast and accurate.

\begin{figure*}
\centering
\includegraphics[scale=1]{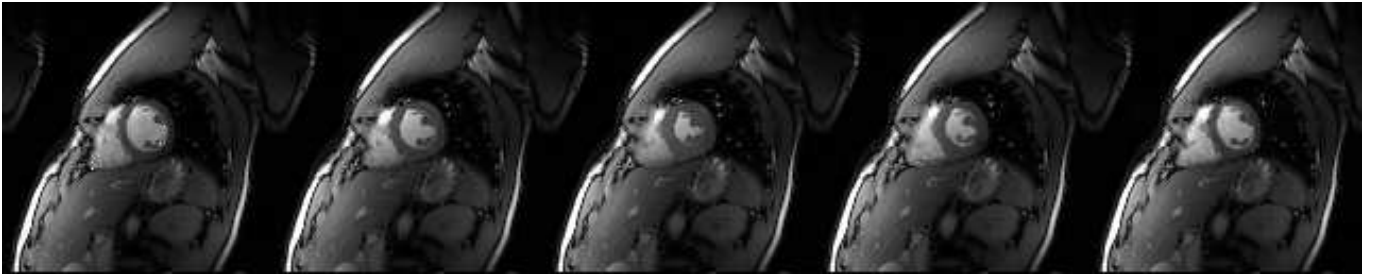}
\caption{Frames 1, 4, 7, 10, 12 of the cine dataset}%
\label{fig:FullSampled}%
\end{figure*}

\subsection{Fast Computation of $(\mu\I + \H'\H)^{-1}$ for Cartesian Sampled Parallel MRI}
\label{sec:fastHconjH}
 Both Algorithms~\ref{alg:admm_synth} and \ref{alg:admm_anlys} rely on fast and efficient computation of $(\mu\I + \H'\H)^{-1}$ especially for a large $\H$. We propose an efficient and exact calculation by making use of the separability of the transfer function among the horizontal and vertical axes in Cartesian sampled Parallel MRI. Deriving $\H'\H$ from (\ref{eqn:H_1}) and (\ref{eqn:H_2}) and using $\F_\h'\F_\h=\I$ we have
\begin{align}
\H'\H &= \left[ \begin{array}{ccc}
\H_1'\H_1 & \cdots & 0 \\
\vdots & \ddots & \vdots\\
0 & \cdots & \H_{n_t}'\H_{n_t}
\end{array} \right] \\
\H_t'\H_t &= \sum\limits_{c=1}^{n_c} \H_{t,c}'\H_{t,c} \\
				&= \sum\limits_{c=1}^{n_c} \S_c'\tilde{\F}_{\v,t}'\F_\h'\F_\h\tilde{\F}_{\v,t}\S_c \\
               &= \sum\limits_{c=1}^{n_c} [(\s_c\s_c') \odot \tilde{\F}_{\v,t}'\tilde{\F}_{\v,t}] \\
               &= (\tilde{\F}_{\v,t}'\tilde{\F}_{\v,t}) \odot \sum\limits_{c=1}^{n_c} (\s_c\s_c') \label{eqn:HconjH}
\end{align}
Note from (\ref{eqn:HconjH}) that $\H'\H$ is disjoint along the horizontal axis as well as the time axis. Therefore it is possible to have singular value decomposition (SVD) of $\H'\H=\U\L\U'$ as $n_h\times n_t$ separate SVDs of $n_v\times n_v$ matrices such that 
\begin{align}
&\U = \left[ \begin{array}{ccc}
\U_1 & \cdots & 0 \\
\vdots & \ddots & \vdots\\
0 & \cdots & \U_{n_t}
\end{array} \right],
\U_t = \left[ \begin{array}{ccc}
\U_{\v,1,t} & \cdots & 0 \\
\vdots & \ddots & \vdots\\
0 & \cdots & \U_{\v,n_h,t}
\end{array} \right],\\
&\L = \left[ \begin{array}{ccc}
\L_1 & \cdots & 0 \\
\vdots & \ddots & \vdots\\
0 & \cdots & \L_{n_t}
\end{array} \right],
\L_t = \left[ \begin{array}{ccc}
\L_{\v,1,t} & \cdots & 0 \\
\vdots & \ddots & \vdots\\
0 & \cdots & \L_{\v,n_h,t}
\end{array} \right],\\
\label{eqn:separate_x}
&\U_{\v,i,t}\L_{\v,i,t}\U_{\v,i,t}' = (\tilde{\F}_{\v,i,t}'\tilde{\F}_{\v,i,t}) \odot \sum\limits_{c=1}^{n_c} (\s_{\v,i,c}\s_{\v,i,c}') 
\end{align}
which is significantly easier to compute than non-separable case. Assuming we have the SVD of $\H'\H=\U\L\U'$ where $\L$ is a diagonal matrix with eigenvalues $\{e_k\}$ on main diagonal and $\U\U'=\U'\U=\I$,
\begin{align}
(\mu\I + \H'\H)^{-1} &= (\mu\I + \U\L\U')^{-1} = \U(\mu\I + \L)^{-1}\U'\\
								  \label{eqn:eigen_mu}
                                 &= \U \bar{\L}_{\mu} \U'
\end{align}
with $\bar{\L}_{\mu}$ a diagonal matrix with the diagonal elements $\left\lbrace \dfrac{1}{ e_k + \mu} \right\rbrace$ where $e_k$ are the eigenvalues of $\L$. In addition to being significantly faster, the computation of SVD and $(\mu\I + \H'\H)^{-1}$ can easily be implemented in parallel since the equations are disjoint for each column of $\x$ at time $t$, $\x_{\v,i,t} $ (\ref{eqn:separate_x}). 

\section{Experimental Results}

Experiments are conducted on a 2D cardiac MRI dataset acquired on a 3T Siemens Trio scanner using a 32-coil matrix body array. Fully sampled data were acquired using a $128\times128$ matrix (FOV = $320\times320$ mm) and 22 temporal frames. The fully sampled data is then retrospectively undersampled by a factor of 8 using a different variable density random undersampling along vertical axis for each time point as shown in Figure~\ref{fig:Subsampling}. The data is normalized so that the reconstruction of the fully sampled data leads to images with maximum intensity of 1. The sample frames from reconstruction of the fully sampled data can be seen in Figure~\ref{fig:FullSampled}. The subsampled data is reconstructed with Algorithm~\ref{alg:admm_synth} and Algorithm~\ref{alg:admm_anlys} using the SVD method to compute the term $(\mu\I + \H'\H)^{-1}$ and setting temporal DFT as $\Ps$ and temporal TV as $\R$ respectively. The same $\lambda$ value that results in good quality reconstruction is used for all the tested algorithms ($\lambda=0.002$).

The reconstruction performance with formulations in (\ref{eqn:anL1}) and (\ref{eqn:anL1_3}) are also presented as P1($n$) and P2 respectively where $n$ represents the number of conjugate gradient iterations used for P1. The DFT is selected mainly for the purpose of comparing Algorithm~\ref{alg:admm_synth} to P1, and P2 for which the analysis and synthesis prior problems become equivalent and the drawbacks of using CG steps are more apparent. SALSA algorithm with the proposed fast computation of $(\mu\I + \H'\H)^{-1}$ is used for TV regularization only, due to Algorithm 1 and SALSA also being equivalent for orthonormal DFT regularization. The reconstruction with algorithms FISTA (for DFT) and MFISTA (for TV) are provided for comparison to non-ADMM methods. $\mu_i$ for each algorithm is selected empirically through simulations to yield for fastest convergence of each algorithm.  In our experiments, selecting $\mu$ in Algorithm~\ref{alg:admm_synth} (and $\mu_2$ for Algorithm~\ref{alg:admm_anlys}) as 0.06 and the ratio $\frac{\mu_2}{\mu_1}$ in Algorithm~\ref{alg:admm_anlys} as 0.5 or 0.25 resulted in sufficiently fast convergence. The inner iterations for MFISTA algorithm as well as for SALSA are similarly selected as 10.

\begin{figure}[t]
\centering
\subfloat[][\label{fig:dft_cpu}CPU]{
\includegraphics[scale=.5]{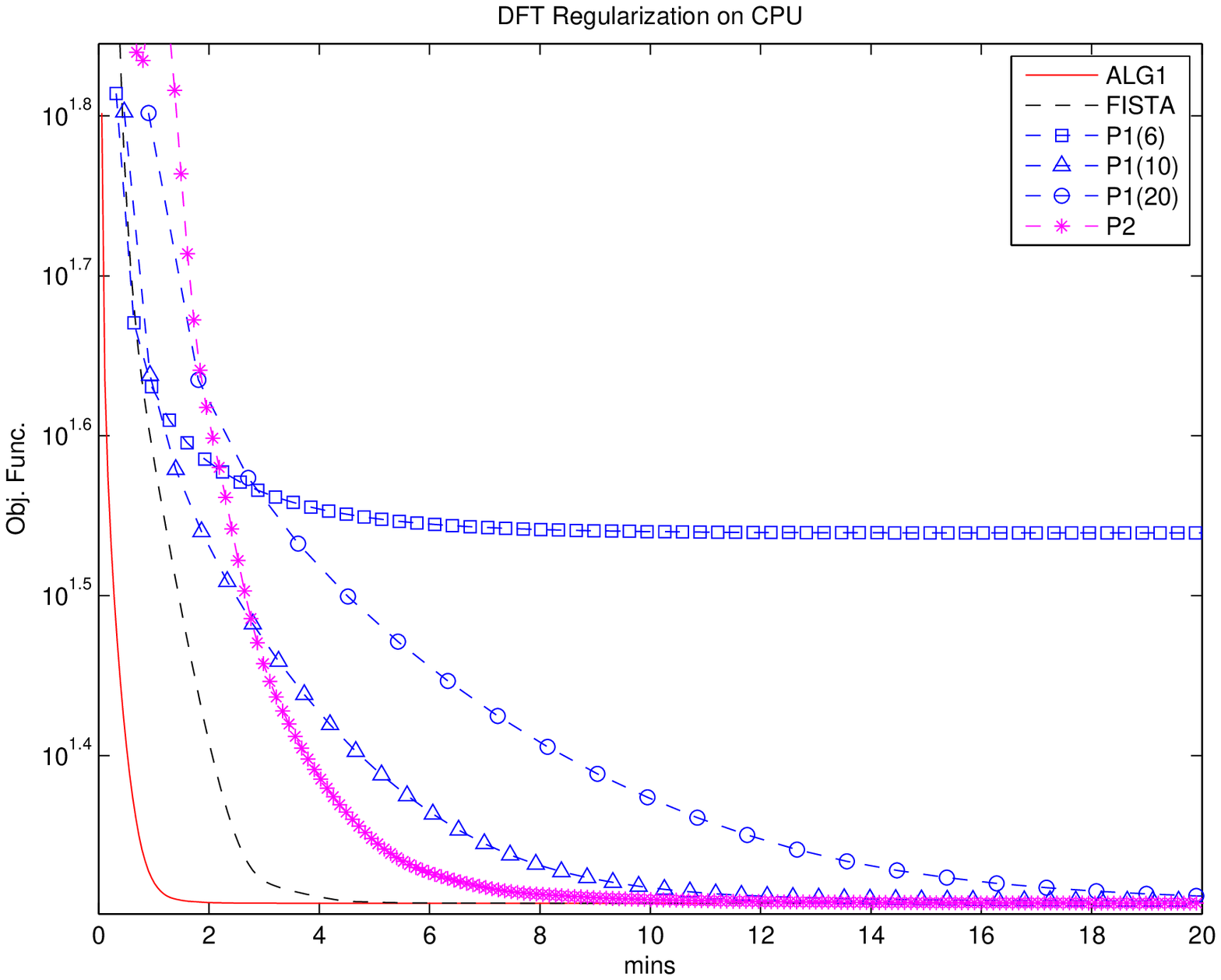}
}%
\qquad
\subfloat[][\label{fig:dft_gpu}GPU]{
\includegraphics[scale=.5]{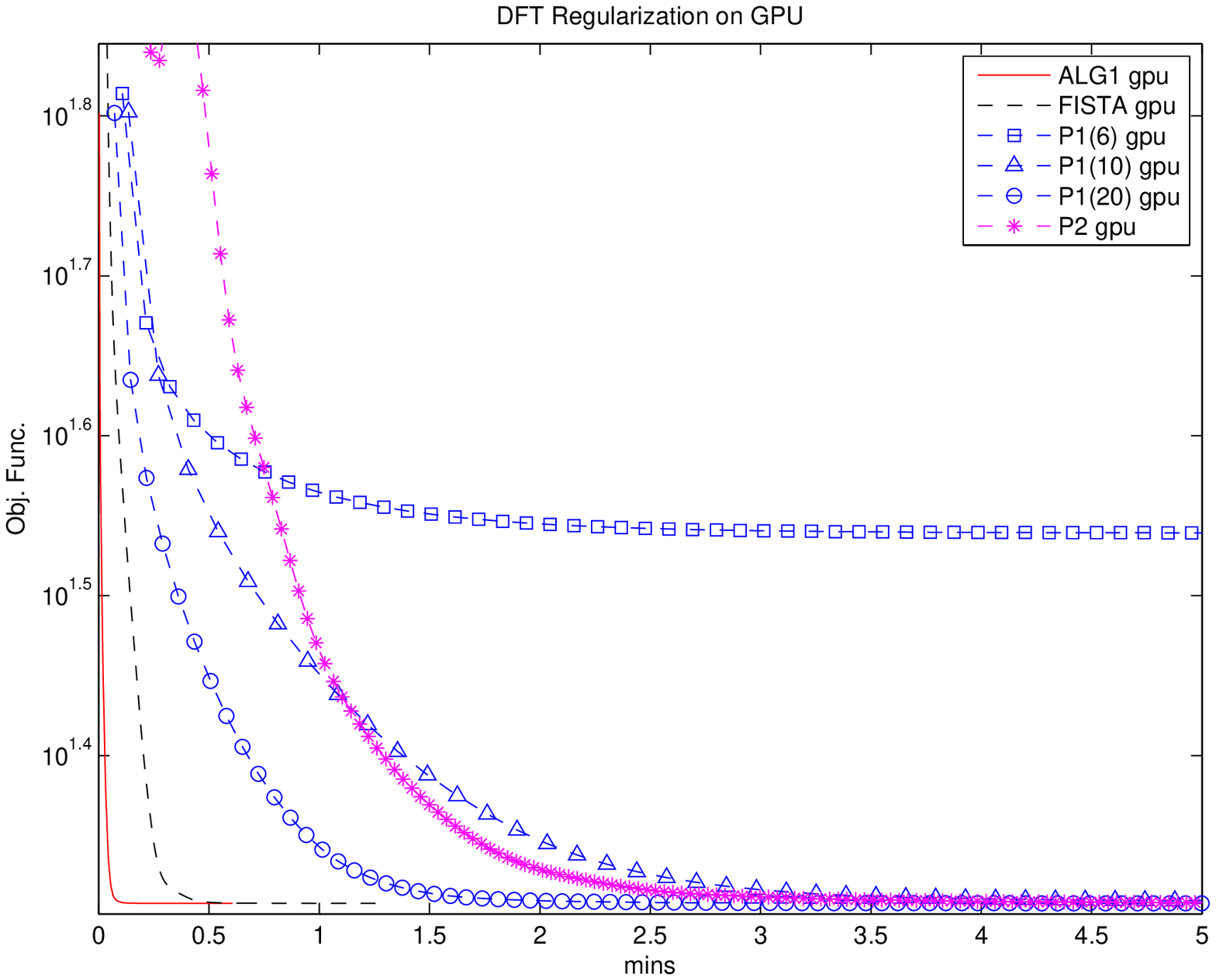}
}
\caption{The objective function for each algorithm with temporal DFT regularization}%
\label{fig:dft_J}%
\end{figure}


All algorithms are implemented in MATLAB$^\copyright$ R2010b \cite{matlab} by the authors and the simulations are performed on a  MacPro5,1 computer with 2.8 GHz quad-core Intel Xeon central processing unit (CPU), 6 gigabytes of memory and MacOSX 10.6.8 operating system. Apart from the simulations on the CPU, each algorithm is also accelerated on GPU with the help of JACKET$^\copyright$ 2.0 toolbox for MATLAB \cite{jacket} and NVidia Quadro FX 4800 graphics card with 1.5 gigabytes of memory and 192 CUDA processor cores to demonstrate how much further each algorithm can be accelerated with the help of parallel processing. JACKET$^\copyright$ 2.0 toolbox provides built in functions to perform almost all operations on GPU which can be as simple as array or matrix multiplication or more complicated such as Fast Fourier Transform or SVD. It also provides the option to perform loops with independent iterations in parallel therefore enabling parallel processing of sequential tasks. All the tested algorithms are implemented in order to utilize GPU as efficient as possible, keeping the steps of the algorithms that does not benefit from GPU acceleration on CPU in order to maximize available memory in the graphics card. Each simulation is initialized with $\x_{initial}=\H'\y$ and performed with 200 iterations for 10 times on minimum operating system load and the fastest among the 10 is reported for the time measure in the results. 

\begin{figure}
\centering
\subfloat[][\label{fig:bar_dft}Temporal DFT Regularization]{
\includegraphics[scale=.5]{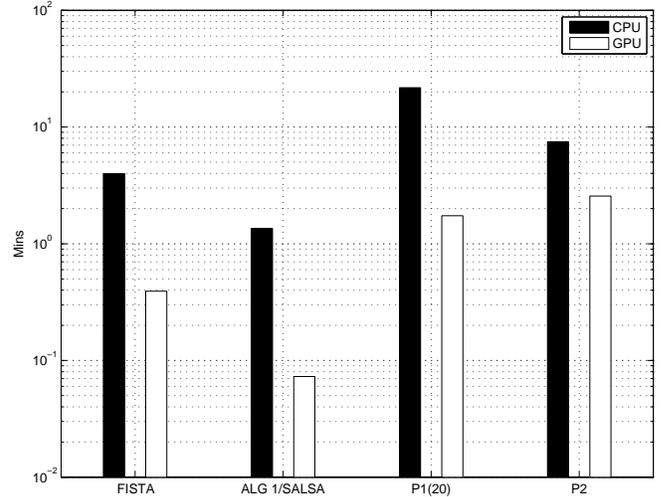}
}%
\qquad
\subfloat[][\label{fig:bar_tv}Temporal TV Regularization]{
\includegraphics[scale=.5]{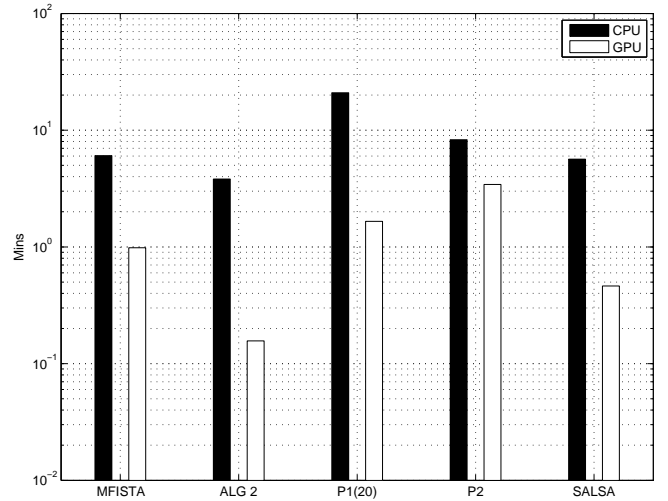}
}
\caption{Comparison of different algorithms in terms of time spent to reach $\Delta(k)=0.001$ defined in (\ref{eqn:r_ratio})}%
\label{fig:bar}%
\end{figure}

The convergence speed of FISTA, Algorithm 1 and P1 for temporal DFT regularization can be seen in Figure~\ref{fig:dft_J}. For an orthonormal transform, Algorithm 1 and P1 have basically the same update on the reconstruction at each iteration, however the difference in their speed is significant mainly due to CG iterations vs SVD based computation. For a severely ill conditioned and large $\H$ as in the experiment, the inaccuracy of CG iterations results in a major impact on the speed of the algorithm. Even when the algorithms are accelerated on GPU, P1 is unable to achieve the speed of CPU implementations of Algorithm 1 or FISTA as observed in Figure~\ref{fig:bar_dft}. P2 algorithm perform better than P1 for CPU simulations whereas P1 is accelerated with GPU implementation much better than P2. This is expected since execution of each step of P2 algorithm is already fast and cannot be accelerated by GPU implementation as much as the other algorithms therefore the slow convergence of P2 algorithm becomes apparent when computational drawbacks are reduced by GPU computation.

\begin{figure}
\centering
\subfloat[][\label{fig:tv_cpu}CPU]{
\includegraphics[scale=.5]{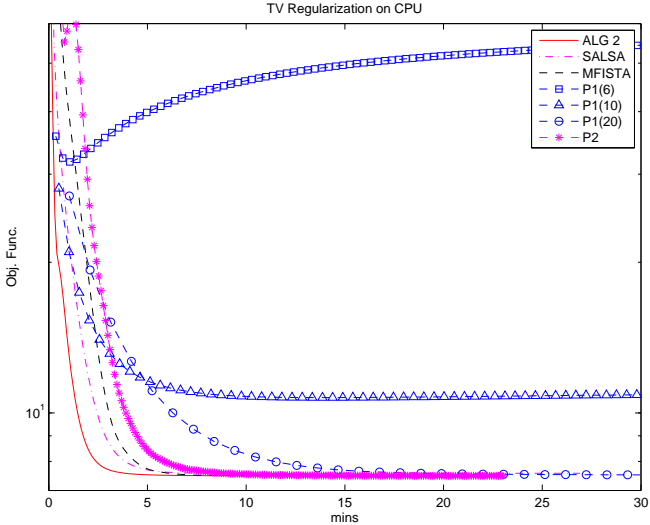}
}%
\qquad
\subfloat[][\label{fig:tv_gpu}GPU]{
\includegraphics[scale=.5]{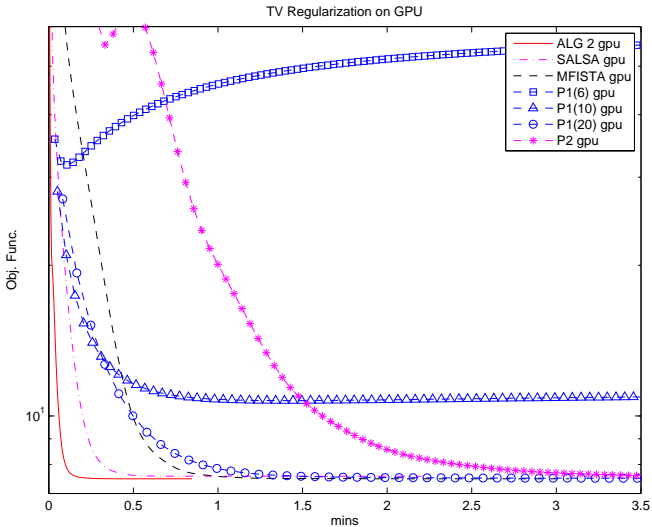}
}
\caption{The objective function for each algorithm with temporal TV regularization}%
\label{fig:tv_J}%
\end{figure}

In our study we use the ratio of reduction in the objective function as the convergence criterion which is defined as
\begin{equation}
\label{eqn:r_ratio}
\Delta(k) \triangleq \dfrac{J(k-1)-J(k)}{J(k)}
\end{equation}
where $J(k)$ is the value of the minimized objective function at iteration $k$. The time it takes for each algorithm to reach $\Delta(k)=0.001$ is shown in Figure~\ref{fig:bar_dft}. It can be seen from Figure~\ref{fig:bar_dft} that Algorithm~\ref{alg:admm_synth} with SVD decomposition is already fastest when executed on CPU, and is accelerated multiple times more than the other algorithms when GPU is utilized. 


For the analysis prior formulation with temporal TV, all the algorithms take longer time compared to DFT regularization case due to $\R$ being ill conditioned as well as $\H$. The performance of the P1 algorithm can be seen to be affected by this the most since the term $(\mu\R'\R + \H'\H)^{-1}$ becomes much harder to estimate. It can be seen from Figure~\ref{fig:tv_J} that the algorithm even fails to converge unless the number of CG iterations are kept higher than 10. Although the GPU acceleration improves the performance, the results are far from the performance of MFISTA or Algorithm 2.  The speed of P2 algorithm is close to P1 for CPU simulation but falls behind during simulations with GPU similar to the case with DFT regularization. SALSA performs better than all other compared algorithms but still falls behind Algorithm~\ref{alg:admm_anlys} due to having multiple iterations of Chambolle's Algorithm for TV minimization at every step. The converged objective function value is also slightly higher than MFISTA and Algorithm~\ref{alg:admm_anlys} due to inaccuracy of the TV estimation with few iterations. The convergence speed of the algorithms can be observed in Figure~\ref{fig:bar_tv}. 

It can be observed from Figures~\ref{fig:dft_J} and \ref{fig:tv_J} that the performance of both P1 and P2 are worse than FISTA and MFISTA unlike the reported performance in \cite{Ramani2011}. The high undersampling ratio in dynamic MRI  applications lead to a challenging reconstruction problem with slow convergence unless the optimization algorithms are fast and accurate, and P1 and P2 are severely affected by $\H$. Both FISTA and MFISTA have computationally simple steps and for a large $\H$ the main computational complexity lies within calculation of $\H'\H$, which can be efficiently calculated thanks to the separability shown in (\ref{eqn:HconjH}). As a result the speed of FISTA and MFISTA are improved much more significantly than P1 and P2 especially when implemented in GPU. The computational speed of these algorithms is therefore comparable to Algorithm~\ref{alg:admm_synth} and \ref{alg:admm_anlys} but overall speed is slower due to faster iterations of ADMM formulation. 

It should be noted that although the presented algorithms are faster, their implementation require a large memory due to simultaneously storing and using the SVD of $\H'\H$. This is not a significant drawback since systems with high memory and computational power are available in commercial configurations. The optimization can also be separated along the horizontal axis when the temporal regularization is used, in order to perform the reconstruction sequentially to reduce the memory requirement if necessary, although this would reduce the acceleration gain. The calculation of the initial SVD of $\H'\H$ is not included in any of the presented simulation results since it can be calculated independently along the horizontal and temporal axis in parallel resulting in only a small computational overhead especially when subsampling ratio is high and $\H'\H$ has a lot of zero eigenvalues.

\section{Conclusions}

In this paper two new ADMM based algorithms are presented to solve the synthesis and analysis prior regularization problem for compressed sensing in parallel MRI. The proposed algorithms make use of the ADMM formulation to provide faster convergence rate than other state of the algorithms such as FISTA, MFISTA, SALSA or traditional Split-Bregman Method. A computational method to improve the speed of the proposed algorithms is also presented which makes use of the separability of the transfer function in Cartesian sampled MRI, leading to faster convergence than other ADMM based methods. The performance of the proposed algorithms are shown to be significantly faster than other methods for a highly undersampled dynamic MRI application and the performance gap with the other algorithms is shown to increase when a parallel processing system is utilized such as GPUs. 

As a drawback of utilizing singular value decomposition of the transfer function to gain speed, implementation of presented algorithms require more memory than other methods and this requirement would increase with larger images or volumes. Although this does not present a problem for systems with large memory, it can still be overcome if needed by reconstructing the signal sequentially along the horizontal axis. The acceleration factor can easily be multiplied with the use of more advanced GPUs and better implementation on faster languages such as C/C++.

\section{Acknowledgements}
The authors would like to thank Li Feng, Daniel Kim, Ricardo Otazo and Daniel K. Sodickson from NYU Medical Center for their help and support as well as for providing the cardiac MRI dataset. We also thank Yilin Song for his assistance in performing the simulations.




%
%
%
\bibliographystyle{IEEEtran_nourl}
\bibliography{libupdate.bib}

\end{document}

%% file: packages.tex
\ifCLASSINFOpdf
  \usepackage[pdftex]{graphicx}
\else
  \usepackage[dvips]{graphicx}
\fi
\usepackage[cmex10]{amsmath}
\usepackage[caption=false,font=footnotesize,justification=centerlast,singlelinecheck=false]{subfig}
\usepackage{fixltx2e}

\usepackage{epstopdf}
\usepackage{booktabs}
\usepackage{dcolumn}
\usepackage{amsfonts}
\usepackage{amssymb}
\usepackage{algorithm}
\usepackage{algpseudocode}
\usepackage{pstricks}